\documentclass[%
 reprint,
superscriptaddress,
 amsmath,amssymb,
 aps,
]{revtex4-2}

\usepackage{graphicx}
\usepackage{dcolumn}
\usepackage{bm}

\DeclareMathOperator*{\argmax}{arg\,max}
\DeclareMathOperator*{\sgn}{sgn}
\newcommand{\bea}{\begin{eqnarray}}
\newcommand{\eea}{\end{eqnarray}}

\begin{document}

\preprint{APS/123-QED}

\title{Detection of the onset of yielding and creep failure from Digital Image Correlation}

\author{Tero M\"{a}kinen}
 \email{Corresponding author\\tero.j.makinen@aalto.fi}
\affiliation{NOMATEN Centre of Excellence, National Centre for Nuclear Research, ul. A. Soltana 7, 05-400  Otwock-\'{S}wierk, Poland}
\affiliation{%
 Department of Applied Physics, Aalto University,
 P.O.  Box 11100, 00076 Aalto, Espoo, Finland
}%
\author{Agata Zaborowska}
\affiliation{NOMATEN Centre of Excellence, National Centre for Nuclear Research, ul. A. Soltana 7, 05-400  Otwock-\'{S}wierk, Poland}
\author{Ma\l{}gorzata Frelek-Kozak}
\affiliation{NOMATEN Centre of Excellence, National Centre for Nuclear Research, ul. A. Soltana 7, 05-400  Otwock-\'{S}wierk, Poland}
\author{Iwona J\'{o}\'{z}wik}
\affiliation{NOMATEN Centre of Excellence, National Centre for Nuclear Research, ul. A. Soltana 7, 05-400  Otwock-\'{S}wierk, Poland}
\author{\L{}ukasz Kurpaska}
\affiliation{NOMATEN Centre of Excellence, National Centre for Nuclear Research, ul. A. Soltana 7, 05-400  Otwock-\'{S}wierk, Poland}
\author{Stefanos Papanikolaou}
\affiliation{NOMATEN Centre of Excellence, National Centre for Nuclear Research, ul. A. Soltana 7, 05-400  Otwock-\'{S}wierk, Poland}
\author{Mikko J. Alava}%
\affiliation{NOMATEN Centre of Excellence, National Centre for Nuclear Research, ul. A. Soltana 7, 05-400  Otwock-\'{S}wierk, Poland}
\affiliation{%
 Department of Applied Physics, Aalto University,
 P.O.  Box 11100, 00076 Aalto, Espoo, Finland
}%

\date{\today}

\begin{abstract}
There is a multitude of applications where structural materials would be desired to be non-destructively evaluated, while in a component, for plasticity and failure characteristics. In this way, safety and resilience features can be significantly improved. Nevertheless, while failure can be visible through cracks, plasticity is commonly invisible and highly microstructure-dependent.
Here, we show that an equation-free method based on principal component analysis is capable of detecting yielding and tertiary creep onset, directly  from strain fields that are obtained by Digital Image Correlation, applicable on components, continuously and non-destructively. We demonstrate the applicability of the method for yielding of Ni-based Haynes~230 metal alloy polycrystalline samples, which are also, characterized through electron microscopy and also, benchmarked using continuum polycrystalline plasticity modeling.  Also, we successfully apply this method to yielding during uniaxial tension of Hastelloy~X polycrystalline samples, and also to the onset of tertiary creep in quasibrittle fiber composites under uniaxial tension. 
We conclude that there are key features in the spatiotemporal fluctuations of local strain fields that can be used to infer mechanical properties.
\end{abstract}

\maketitle


\section{Introduction}

Under small stresses, materials deform in a linear, elastic, manner. With higher stresses the deformation response deviates from the linear one, which can be due to many different reasons e.g. non-linear elasticity, damage, or plasticity. Thus the methods of prediction for such behavior also become highly system dependent.
When the stresses are applied for extended periods of time, slow time-dependent deformation occurs -- the material creeps.
The creep case is more complex than yielding due to e.g. history effects from damage accumulation and plastic deformation buildup as well as the resulting stress redistribution.
An equation-free prediction scheme~\cite{papanikolaou2020microstructural, frydrych2021materials,
biswas2020prediction,liu2020predicting,wang2021machine,wang2022high} for these phenomena would not only be important in practical applications, but also for the fundamental physics involved: What are the universal features that can be exploited?\\

Yielding represents a transition from an elastic state to a plastic one.
There exists several engineering definitions of yielding~\cite{dieter1976mechanical}, such as the proportionality limit or the offset yield point. The former is the point when the deviation of the stress-strain behavior from the Hookean behavior exceeds a predetermined threshold (for example 1~\%) and the latter is simply the stress at a predetermined point, commonly at 0.2~\% engineering strain (denoted usually as~$\sigma_{0.2\%}$). 
The important question is then: are these engineering definitions of yielding good enough?
To at least avoid the problem of predetermined thresholds and constant points we define here the yield strain to be the maximum of the second derivative of stress with respect to strain~\cite{papanikolaou2021direct}
\begin{equation} \label{eq:yield}
    \epsilon_{y} = \argmax \frac{\partial^2 \sigma}{\partial \epsilon^2}
\end{equation}
which nicely captures the point of maximum curvature.

The materials we have used for yield testing are two nickel based superalloys. They have been chosen due to their excellent mechanical properties~\cite{darolia2019development,
pollock2006nickel,pollock2016alloy,murray2020defect} which make them important for range of practical applications, in particular high-temperature applications.\\

However in many practical applications the loading of materials is not through a constantly increasing stress. Instead the load is static, leading to time-dependent creep behavior.
Moreover, in creep defining the onset of failure is even more difficult and the creep characteristics of different materials can vary. The common thing is that creep failure is preceded by an acceleration of strain accumulation (tertiary creep)~\cite{nechad2005creep, leocmach2014creep, koivisto2016predicting, makinen2020scale}, so a natural definition for the onset of creep failure is the point of minimum strain rate. 
This can also be thought of as a transition from one state to another -- from a decelerating strain evolution to an accelerating one.
The ideal thing for applications would then be a touch-free non-destructive testing (NDT) method for determining this point of creep failure onset using easily acquirable data.

As a test material for creep we have used a quasibrittle disordered material -- paper.
The creep behavior of paper is fairly well known~\cite{alavaniskanen2006,koivisto2016predicting, makinen2020scale} and can be collapsed to a single master curve. The behavior divides into three phases: primary, logarithmic, and tertiary creep. Primary and logarithmic creep regimes are characterized by a power-law decrease of the strain rate until a strain rate minimum is reached. 
After this minimum -- in the tertiary creep regime, the strain rate increases rapidly which finally leads to the failure of the sample.
In paper and the geometry used in these experiments the strain rate minimum has been observed to occur at $t = 0.83 \times t_c$~\cite{koivisto2016predicting} where $t_c$ denotes the time of failure of the sample. This statistical relation is known as the Monkman--Grant relation~\cite{monkman1956american} and provides a method of predicting the sample failure time from the strain rate minimum. However the real-time determination of the minimum from a noisy strain rate signal is a difficult task.\\

\begin{figure}[t!]
    \centering
    \includegraphics[width=\columnwidth]{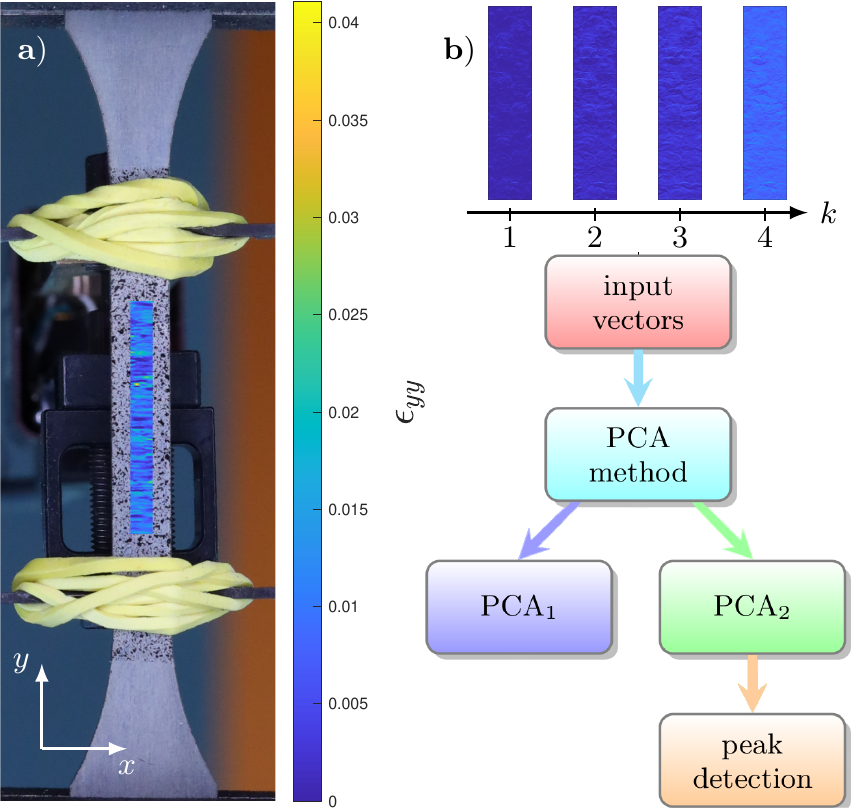}
    \caption{\textbf{a}) An image of the sample during a tensile yielding test, showing the applied speckle pattern and the extensometer measuring the global strain in the sample. The local strain field $\epsilon_{yy}$ obtained from DIC is superimposed on the sample.
    \textbf{b}) A schematic representation of the PCA based detection method used. The method takes as an input a set of strain fields at timesteps $k$, performs PCA and outputs two PCA components. The second PCA component has a clear peak corresponding to the maximum fluctuations in the system.}
    \label{fig:fig1}
\end{figure}

Digital Image Correlation~(DIC)~\cite{hild2006digital, pan2009two, jones2018dic} has been a popular NDT technique for a fairly long time. Data for DIC can be acquired easily using regular cameras and a plethora of easy-to-use software~\cite{cintron2008strain, turner2015digital, blaber2015ncorr, yang2019augmented} exists for DIC computations. However the strain maps obtained using DIC (for an example see Fig.~\ref{fig:fig1}a) are commonly used only to study e.g. the localization phenomena related to deformation \cite{orozco2017magnesium,bourdin2018measurements,stinville2020direct,latypov2021insight,orozco2021high,vermeij2022plasticity} or the shape of the strain field in a complex geometry.
The information "hidden" in the DIC images is underutilized and exploiting it in a meaningful way would be a major goal for materials informatics.\\

To capture these transitions to plasticity dominated states we provide in this paper a solution based on Principal Component Analysis~(PCA)~\cite{wang2016discovering, hu2017discovering, wetzel2017unsupervised, wang2017machine, ruscher2019correlations, quinn2019visualizing, salmenjoki2020probing}. The main idea of the method is to probe the spatial fluctuations in the local strain fields, as these are a characteristic feature of plasticity. The detection method takes as an input a set of strain fields (computed using DIC) and outputs two principal components which show the transition between two states and a uniquely defined transition point (point of maximum fluctuations, seen as a peak in one of the components). A schematic of the method can be seen in Fig.~\ref{fig:fig1}b and
the method has previously been shown to work well with simulated strain fields \cite{papanikolaou2021direct} mimicking the fields obtained using DIC.
In this paper the method is used in detecting the yielding of two commercial metal alloys during monotonic tensile loading of the sample. Additionally the method is used to detect the onset of tertiary creep in paper from similar images.

\section{Methods} 

\subsection{Experiments}

\begin{table}[t]
    \centering
     \begin{tabular}{c c c}
       \hline
        & \textbf{Haynes~230} & \textbf{Hastelloy~X}  \\
        & [wt. \%] & [wt. \%] \\
        \hline
        Al & 0.37 & 0.11 \\
        B & 0.004 & $<$0.002 \\
        C & 0.1 & 0.070 \\
        Co & 0.2 & 1.22 \\
        Cr & 21.87 & 21.27 \\
        Cu & 0.03 & 0.09 \\
        Fe & 1.23 & 18.83 \\
        Mn & 0.50 & 0.64 \\
        Mo & 1.46 & 8.29 \\
        Ni & Bal. & Bal. \\
        P & 0.007 & 0.015 \\
        Si & 0.31 & 0.24 \\
        W & 14.27 & 0.52 \\
        S & $<$0.002 & $<$0.002 \\
        Ti & $<$0.01 & $<$0.01 \\
        \hline
    \end{tabular}
    \caption{Chemical composition of the two metal alloys used in the yielding experiments as weight percentages provided by the sample manufacturer Haynes International Company.}
    \label{tab:comp}
\end{table}

To test the yielding of metal alloys, tensile tests were performed on the Hastelloy~X and Haynes~230 alloys provided by Haynes International Company which were cold-rolled and annealed. The chemical composition of the alloys can be seen in Table~\ref{tab:comp}.
Flat dogbone-shaped tensile specimens (in accordance with ISO~6892-1 standard) with gauge length of 30~mm, width of 5~mm (Haynes~230) or 4~mm (Hastelloy~X), and thickness of 1.02~mm (Haynes~230) or 1.3~mm (Hastelloy~X) were cut by Wire Electrical Discharge Machining (WEDM) method using Robofil~200 machine. 
The thickness of the samples was the same as the thickness of the provided sheets. 
Three samples of each alloy were tested at room temperature using the Instron~8501 servo-hydraulic system. Direct strain measurement was performed by axial clip-on dynamic extensometer with original gauge length of 25~mm provided by Instron. 
Tensile experiments were performed in accordance with both ISO~6892-1 and ASTM~E8 standards. To ensure appropriate alignment of the loading string, the preliminary stress of 15~MPa was applied before each measurement started. This value did not exceed 5~\% of the yield strength of each material. 
Up to the yielding point, testing speed was defined in stress rate and set for 10~MPa/s. 
After the displacement reached the value of~0.5~mm, the test rate was changed to an estimated strain rate of 0.001~s${}^{-1}$.\\

The microstuctural characterization of the Haynes~230 and Hastelloy~X samples was conducted using ThermoFisher Scientific Helios~5~UX field emission gun high-resolution scanning electron microscope~(HR-SEM) equipped with an EDAX Velocity Pro electron backscatter diffraction~(EBSD) system. The EBSD measurements were conducted using an accelerating voltage of 20~keV and a beam current of 3.2~nA. The EBSD grain orientation inverse pole figure~(IPF)~Z~maps (shown in Fig.~\ref{fig:figEbsd}a,b) were acquired using a step size of 300~nm. The maps were analyzed using the EDAX OIM Analysis 8 software to remove misindexed points by requiring the confidence index to exceed~0.1. SEM~images (Fig.~\ref{fig:figEds}a,b) were collected at 5 keV electron beam using in-column detector~(ICE) in secondary electrons~(SE) contrast, and energy dispersive X-ray spectroscopy (EDS) was used to perform mapping of atomic concentration distribution of individual elements (Fig.~\ref{fig:figEds}c-h).

Prior to the measurement the surface of the samples was prepared by using the LectroPol-5 system provided by Struers. Samples of both alloys underwent surface treatment by using mixture of methanol and perchloric acid (for Hastelloy~X) and ethanol with perchloric acid (for Haynes~230). The process was conducted at 10~C until microstructural features were revealed.\\

During loading, the samples are imaged using Canon EOS~R camera with a frequency of 0.5~Hz and illuminated by a ring lamp. For texture a speckle pattern is sprayed on the samples. 
From these images, the DIC calculations were performed using AL-DIC software~\cite{yang2019augmented} 
which calculates the local displacements $\bm{u} = (u, v)$ with respect to the first image using a circular region of interest with a radius of 1.5~mm 
and placing the regions of interest every 25~$\mu$m. 
The area used in the DIC calculations -- the region of interest -- only includes the middle part of the sample, to avoid any boundary effects.
The local strain component in the loading direction is then calculated from these displacements as $\epsilon_{yy} = \frac{\partial v}{\partial y}$ using simple finite difference numerical differentiation.
The size of the resulting strain fields is 100~pixels $\times$ 800~pixels.
An image of the sample and the strain field obtained by DIC can be seen in Fig.~\ref{fig:fig1}a. \\

The experimental details of the paper creep experiments are described in Ref.~\cite{makinen2020scale}. The data used here includes the global strain rate $\dot{\epsilon}$ measured by the tensile testing machine and the local strains in the loading direction $\epsilon_{yy}$ measured at different points in time. The spatial resolution of the images is 100~$\mu$m.\\

\subsection{Detection method}

We start by taking the computed strain fields $\epsilon_{yy}^{(k,i)}$ (where the $i$ index runs over all $V$ spatial points and $k$ denotes the timestep) and normalize them to a matrix $X$ consisting of the normalized (average value of zero and standard deviation of unity) input vectors as rows
\begin{equation}
    X_{(k,i)} = \frac{\epsilon_{yy}^{(k,i)} - \left\langle \epsilon_{yy}^{(k,i)} \right\rangle}{\sqrt{\left\langle \left( \epsilon_{yy}^{(k,i)} \right)^2 \right\rangle - \left\langle \epsilon_{yy}^{(k,i)} \right\rangle^2}}
\end{equation}
where $\langle \cdot \rangle$ denotes a spatial average.
Here we have only considered the $\epsilon_{yy}$ component but naturally one could make other choices, if deemed appropriate for the situation.
We then compute the singular value decomposition
\begin{equation}
    X = U \Sigma W^\mathrm{T}
\end{equation}
where the columns of $U$ are the left singular vectors, the columns of $W$ the right singular vectors, and the diagonal matrix $\Sigma$ contains the singular values of $X$. For each singular value $\sigma_j$ there exists a singular vector $s_j$.

The method captures the fluctuations by computing the eigenvalues $\lambda$ of the covariance matrix of the input vectors
\begin{equation}
    C = \frac{X^\mathrm{T} X}{V-1} = U \left( \frac{\Sigma^2}{V-1} \right) U^\mathrm{T}
\end{equation}
which can clearly be seen to relate to the singular values of $X$ through the relation $\sigma_j = \sqrt{\lambda_j (V-1)}$.

Finally the sorted singular vectors (starting with the largest one) are projected onto the input vectors giving the PCA components
\begin{equation}
    \mathrm{PCA}_j^k = \frac{s_j \cdot X_{k}}{\sqrt{\sigma_j}}
\end{equation}
which should be thought of as just a timeseries $\mathrm{PCA}_j(t)$ for each PCA component.\\

The main idea is that only a few of the components already capture most of the fluctuations present in the dataset. In case of material behavior the natural interpretation of a system described by two components is that the strain can be divided into two contributions: reversible elastic strain and irreversible plastic strain.

\subsection{Simulations}

\begin{figure*}[t!]
    \centering
    \includegraphics[width=\textwidth]{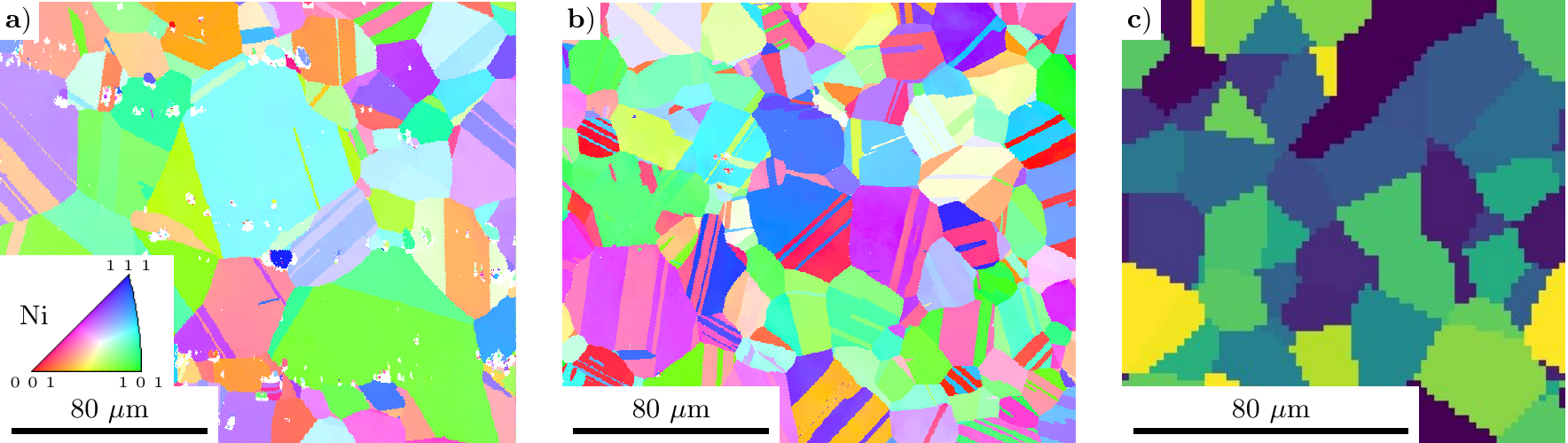}
    \caption{\textbf{a}) An EBSD IPF Z map of the a Haynes~230 sample.
    \textbf{b}) An EBSD IPF Z map of a Hastelloy~X sample (map legend for both samples is presented in the inset of panel~a).
    \textbf{c}) The grain structure used in the Haynes~230 simulation, assuming a pixel's linear dimension to be 2 $\mu$m.}
    \label{fig:figEbsd}
\end{figure*}

For comparison we also simulated a similar uniaxial tensile test and the resulting yielding using a standard crystal plasticity model, with material parameters that match the behavior of the Haynes~230 alloys, following the prescriptions in Ref.~\cite{luccarelli2017finite} that investigated in detail the way to match polycrystalline Haynes~230 data. Also, the grain structure used in the model has been chosen through a Voronoi cell tesselation, which aims to emulate the grain structure seen in EBSD measurements (see Fig.~\ref{fig:figEbsd}a and c), which is characterized by randomly oriented grains with 10-20 $\mu$m linear dimension.

\begin{table}[tb]
    \centering
     \begin{tabular}{c c c}
       \hline
        \textbf{Model Parameters} & \textbf{Symbol} & \textbf{Value}  \\
        \hline
        Dimensions & $L_x$, $L_y$, $L_z$ & 64 px, 64 px, 64 px \\
        & & 128 $\mu$m, 128 $\mu$m, 128 $\mu$m \\
       Elastic stiffness  & $C_{11}$ & 323 GPa \\
       Elastic stiffness & $C_{12}$ & 159.1 GPa \\
       Elastic stiffness & $C_{44}$ & 89.5 GPa \\
       Reference shear rate & $\dot\gamma_0$ &  0.001 s${}^{-1}$  \\
      Rate sensitivity exponent & $m$ & 0.0015  \\
      Slip-Slip Interaction & $h_0$ & $400$ MPa \\
      Slip hardening parameter & $p$ & 2.25 \\
      Saturated shear resistance & $\tau^s$ & 678 MPa \\       
        \hline
    \end{tabular}
    \caption{Model parameters chosen in this work, chosen in accordance to the prescriptions in the thorough study of Haynes~230 polycrystalline alloy samples in Ref.~\cite{luccarelli2017finite}.}
    \label{tab:1}
\end{table}

We  study~\cite{pap2019, asaro2006mechanics, roters2019, papanikolaou2021direct}  tensile loading in the $x$-direction, for 3D polycrystalline samples with sample dimensions in $(x,y,z)$: (64,64,64) {  (3D)},   {  promoting the perspective of investigating a (0.13 mm)${}^3$ sub-mm cubic 3D samples.}. We assume that the linear size of each cubic pixel is 2 $\mu$m, and the chosen Voronoi tesselation (see Fig.~\ref{fig:figEbsd}c) points to grains with linear dimension 10-20 $\mu$m, as in experimental samples~(see Fig.~\ref{fig:figEbsd}a). The crystalline structure of the material is {face-centered cubic (FCC) Aluminum (Al),} with standard stiffness coefficients (see Table~\ref{tab:1}, in reference to the cubic coordinates). The digital surface image collection is assumed to be collected at periodic applied strain intervals, similar to the experimental study. The developed model in this work displays similar yield point with the experimental Haynes~230 sample (see Fig.~\ref{fig:fig3}), and the plastic yielding differences can be attributed to prior processing and microstructural details that lie beyond the purpose of this work and also, Ref.~\cite{luccarelli2017finite}.

The model~\cite{pap2019, roters2019} utilizes the phenomenological crystal plasticity theory, capturing   slip-based macroscale plasticity, with  standard constitutive laws for metals~\cite{asaro2006mechanics}. The model captures in a self-consistent manner the basic physical mechanisms of crystal plasticity, as they take place in most metals, and  it captures finite deformations in a cubic grid~\cite{papanikolaou2021direct}. The model is solved by using a FFT-based spectral method~\cite{pap2019}. The plastic deformation tensor evolves as:
\bea
\dot{F_p} = L_p F_p
\eea
where $L_p=\sum_\alpha \dot\gamma^\alpha s^\alpha\otimes n^\alpha$, and s, n unit vectors on slip direction and slip plane normal, respectively, while $\alpha$ is the slip system index. Total deformation translates in elastic and plastic ones through $F=F^eF^p$. The slip rate $\dot\gamma^\alpha$ is given by~\cite{asaro2006mechanics}, 
\bea
\dot\gamma_\alpha = \dot\gamma_0 \bigg|\frac{\tau^\alpha}{g^\alpha}\bigg|^n \sgn(\tau^\alpha)
\eea
where $\dot\gamma_0$ is the reference shear rate, $\tau^\alpha = S\cdot (s^\alpha \otimes n^\alpha)$ is the resolved shear stress at a slip resistance $g^\alpha$, with~$S=[C]E^\epsilon$ being the Second Piola-Kirchoff stress tensor, n) in this work (inverse of the strain rate sensitivity exponent $m=1/n$) and $g^\alpha$ is the slip resistance for a slip system $\alpha$.  Hardening is provided by: 
\bea
\dot g^\alpha = \sum_{\beta=1}^{12} h_{\alpha\beta} |\dot\gamma^\beta|
\eea
where $h_{\alpha\beta}$ is the hardening matrix, which captures the micromechanical interactions between different slip systems, and the shear resistances asymptotically evolve towards saturation. For more details on the model details, please see associated references~~\cite{pap2019, asaro2006mechanics, roters2019, papanikolaou2021direct}.

\begin{figure}[h!]
    \centering
    \includegraphics[width=\columnwidth]{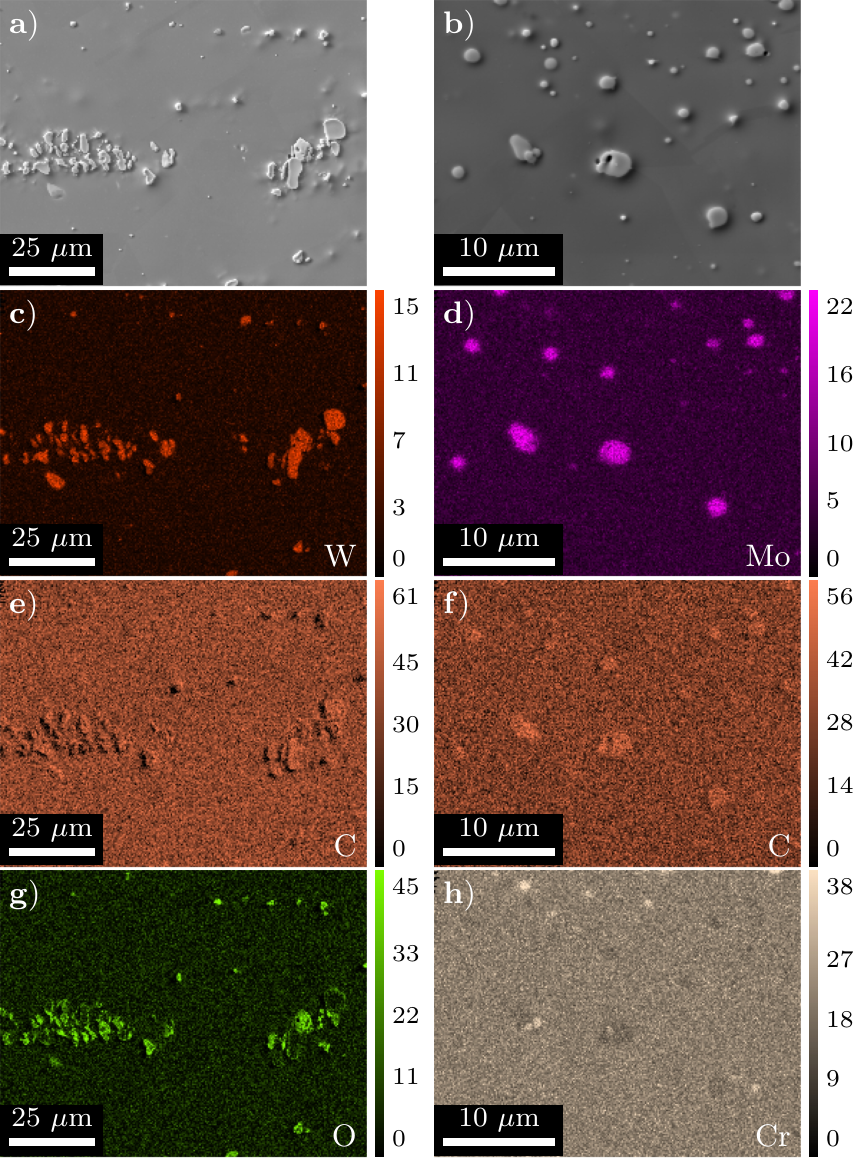}
    \caption{\textbf{a}) SEM image of the Haynes~230 sample.
    \textbf{b}) SEM image of the Hastelloy~X sample.
    \textbf{c}, \textbf{e}, \textbf{g}) EDS atomic concentration maps (in at.~\%) of W, C, and O in the Haynes~230 sample.
    \textbf{d}, \textbf{f}, \textbf{h}) EDS atomic concentration maps (in at.~\%) of Mo, C, and Cr in the Hastelloy~X sample.}
    \label{fig:figEds}
\end{figure}

\begin{figure}[h!]
    \centering
    \includegraphics[width=\columnwidth]{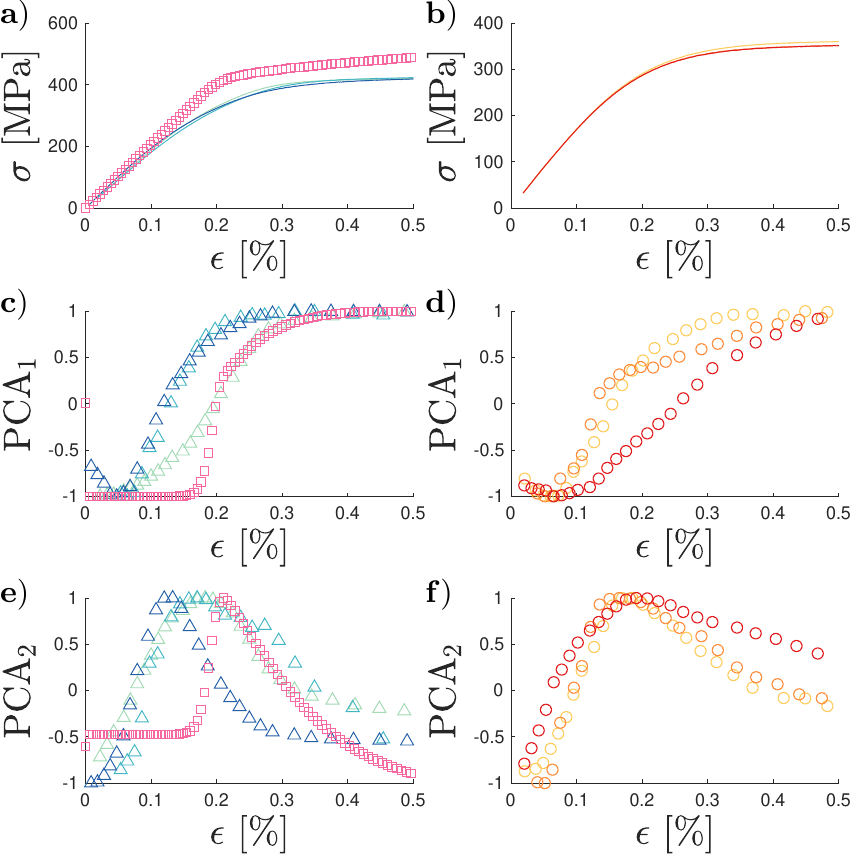}
    \caption{\textbf{a},~\textbf{b}) The stress-strain response of the metal alloys. The three curves represent the three different samples. The purple rectangles in the Haynes~230 plot correspond to the results of the simulation.
    \textbf{c},~\textbf{d}) The first PCA component as a function of the global strain, colors as in panels a, b. A clear transition from a low to a high value can be seen around a point corresponding to the yielding of the sample. The purple rectangles correspond to the simulation.
    \textbf{e},~\textbf{f}) The second PCA component as a function of the global strain, colors as in previous panels. A peak is observed for each curve, around a point corresponding to the yielding of the sample. The purple rectangles correspond to the simulation.}
    \label{fig:fig2}
\end{figure}

\section{Results}

\subsection{Microstructural characterization}

The IPF~Z~maps (Fig.~\ref{fig:figEbsd}a,b) of the Haynes~230 and Hastelloy~X samples show an isotropic grain structure with an average grain area of 31.29~$\mu$m${}^2$ and 16.8~$\mu$m${}^2$, respectively.
The twinning present in the maps is related to the material processing. The samples were cut from cold-rolled and annealed plates, and this process gives rise to deformation twins.

The grain orientation maps and SEM images (Fig.~\ref{fig:figEds}a,b) reveal the presence of precipitates with strongly developed topography structures standing out the matrix surface prepared by electrochemical  polishing, both for Hastelloy~X and Haynes~230 samples. In the SEM images a channelling contrast between the grains of specific orientations along the zone axis can also be observed.
On the basis of comparative analysis performed by SEM, EBSD and EDS techniques (Figs.~\ref{fig:figEbsd} and \ref{fig:figEds}) the precipitates have been identified as inclusions of non-metallic phases responsible for misindexing of the EBSD patterns in specific areas, represented in white in the EBSD orientation maps.
Careful analysis of the EDS data allowed us to identify the precipitates as 
Tungsten carbides and Tungsten oxides in the case of Haynes 230 sample (Fig.~\ref{fig:figEds}c,e,g), and
Molybdenum carbides and Chromium carbides in the case of Hastelloy X sample (Fig.~\ref{fig:figEds}d,f,h).
It should be kept in mind that due to the inhomogenity in the composition of the samples and the accuracy of the quantitative EDS analysis in general, the values of atomic concentrations of individual elements presented in EDS elemental maps must be treated as approximate (with a high level of uncertainty).

\begin{figure}[h!]
    \centering
    \includegraphics[width=\columnwidth]{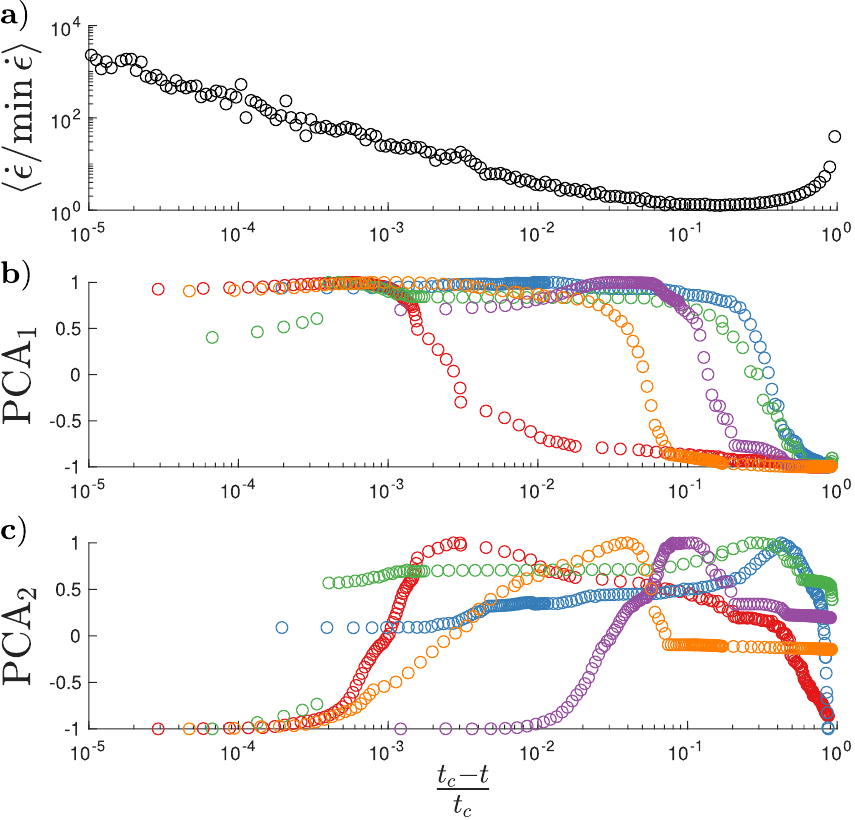}
    \caption{\textbf{a}) The strain rate $\dot{\epsilon}$ normalized by the strain rate minimum $\min \dot{\epsilon}$ averaged over all paper creep experiments as a function of the normalized time before failure $(t_c-t)/t_c$. Note that the time in this plot goes from right to left. Initially the strain rate decreases until it reaches the minimum at $t_{min}$ and after that increases as a power-law towards sample failure at $t_c$.
    \textbf{b}) The first PCA component as a function of the normalized time before failure for five samples. A clear transition from a low to a high value can be seen for each sample around a point corresponding to the onset of strain rate increase (so onset of tertiary creep).
    \textbf{c}) The second PCA component as a function of the normalized time before failure (colors as in panel b). A peak is observed for each sample around a point corresponding to the transition observed in the first PCA component.}
    \label{fig:fig3}
\end{figure}

\begin{figure}[h!]
    \centering
    \includegraphics[width=\columnwidth]{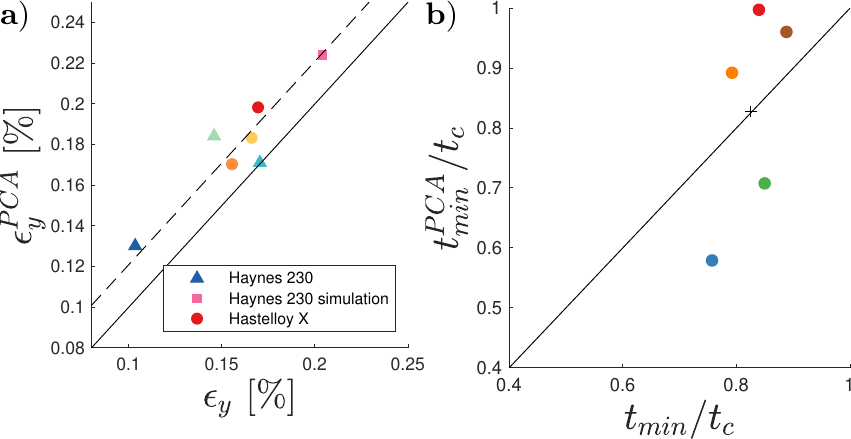}
    \caption{\textbf{a}) The location of the peak in the second PCA component $\epsilon_y^{PCA}$ (Eq.~\ref{eq:prediction}) in the metal alloy experiments plotted against the yield strain $\epsilon_y$ determined from the stress-strain curves (Eq.~\ref{eq:yield}), colors as in Fig.~\ref{fig:fig2}. The black line corresponds to the case $\epsilon_y^{PCA} = \epsilon_y$ and the dashed line has a constant offset of 0.02 \%.
    \textbf{b}) The location of the peak in the second PCA component $t_{min}^{PCA}$ (Eq.~\ref{eq:prediction2}) in the creep experiments plotted against the time of the strain rate minimum $t_{min}$, colors as in Fig.~\ref{fig:fig3}. The black line corresponds to the case $t_{min}^{PCA} = t_{min}$ and the cross marker to the average off all the experiments.}
    \label{fig:fig4}
\end{figure}

\subsection{Detection of yielding}

The measured stress-strain response (seen in Fig.~\ref{fig:fig2}a,b) has been used to determine the yield strain for each metal sample. This is done by finding the maximum of the second derivative of stress with respect to strain (Eq.~\ref{eq:yield})
which then can be compared with the results obtained using the PCA method.
The Haynes~230 samples can be seen to yield at $\epsilon_y = 0.14 \pm 0.03$~\% which corresponds to a yield stress of $\sigma_y = 249 \pm 44$~MPa, and an elastic modulus of $E = 180 \pm 13$~GPa. For the Hastelloy~X samples these numbers are $\epsilon_y = 0.16 \pm 0.01$~\%, $\sigma_y = 254 \pm 8$~MPa, and $E = 156 \pm 2$~GPa.

The simulated stress-strain curve can also be seen in Fig.~\ref{fig:fig1}a. It roughly matches the experimentally observed behavior with an elastic modulus of 200~GPa, but with a much sharper yield point, yielding at 0.20~\% strain. This increased curvature around yielding also increases the yield stress to 408~MPa.\\

The PCA method takes as an input a set of strain fields at different stages of loading, performs Principal Component Analysis and gives out two principal components which we here denote by $\mathrm{PCA}_1$ and $\mathrm{PCA}_2$. Plotting these components as a function of strain (see Fig.~\ref{fig:fig2}c,d,e,f) shows that the first component strongly increases around yielding and after that stays constant and the second component has a clear peak around yielding. For clarity of presentation, the extreme values of the PCA components have here been normalized to positive and negative unity.\\

These observations on the locations of the peaks in the second PCA component motivate the following definition~\cite{papanikolaou2021direct} for the PCA based prediction of yielding. We define the yield strain prediction as the position of the peak
\begin{equation} \label{eq:prediction}
    \epsilon_{y}^{PCA} = \argmax \mathrm{PCA}_2
\end{equation}
and this clearly is close to the yield strain defined from the stress-strain curves using Eq.~\ref{eq:yield}.

\subsection{Detection of creep failure onset}

For the paper creep experiments it is most illustrative to plot the strain rate (normalized by the strain rate minimum) averaged over all experiments $\langle \dot{\epsilon} / \min \dot{\epsilon} \rangle$ as a function of (normalized) time before failure $(t_c-t)/t_c$ as can be seen in Fig.~\ref{fig:fig3}a. Initially the strain rate decreases over time and a minimum is seen on average at $t = 0.83 \times t_c$ or $(t_c-t)/t_c = 0.17$ after which the strain rate increases in a power-law fashion towards the failure of the sample.

Doing the exact same procedure as previously for the strain fields from the paper creep experiments again yields two principal components. Plotting them as a function of the (normalized) time before failure (see Fig.~\ref{fig:fig4}b,c) one can see that similarly to the previous case, the first component increases from a low constant value to a high constant value at a point which seems to correspond to the onset of tertiary creep (so around the strain rate minimum). Again, similarly to the yielding case, the second component has a clear peak around the same point.\\

Similarly to the yielding case we take the peak of the second PCA component to correspond to the onset of tertiary creep (which would correspond to the time of the strain rate minimum $t_{min}$). This then naturally gives the definition
\begin{equation} \label{eq:prediction2}
    t_{min}^{PCA} = \argmax \mathrm{PCA}_2
\end{equation}
analogous to the definition of Eq.~\ref{eq:prediction}.

\begin{figure}[t!]
    \centering
    \includegraphics[width=\columnwidth]{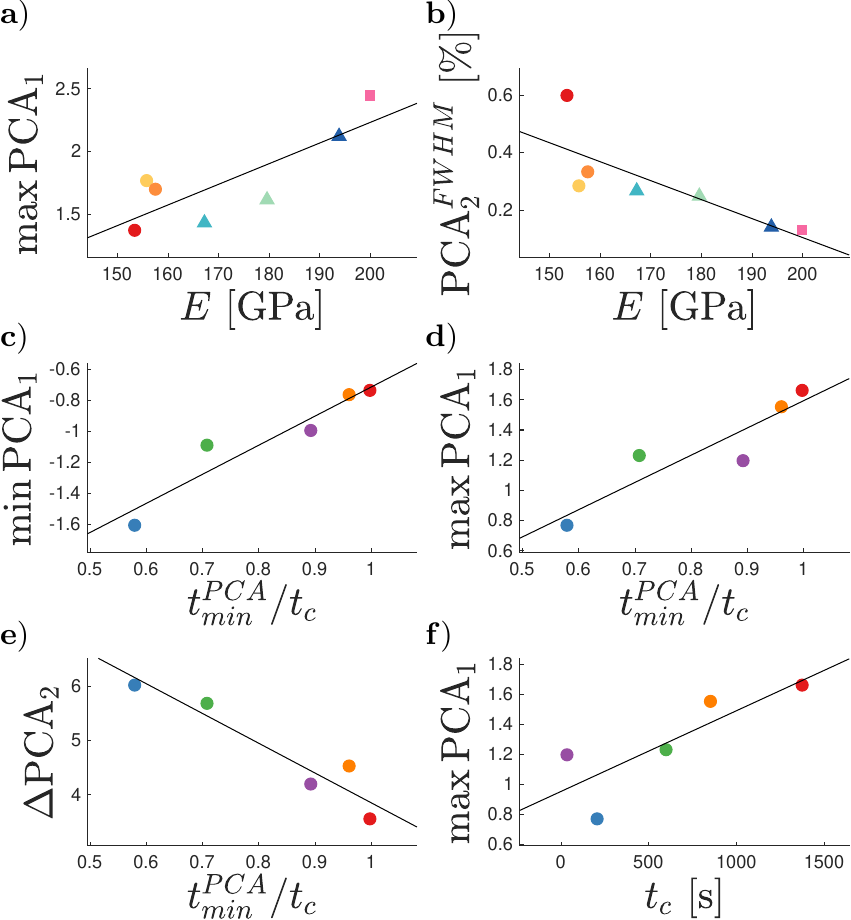}
    \caption{
    The colors and markers for panels a,b as in Figs.~\ref{fig:fig2} and \ref{fig:fig4}, and
    colors for panels c-f as in Figs.~\ref{fig:fig3} and \ref{fig:fig4}.
    \textbf{a}) The maximum value of PCA${}_1$ component plotted against the elastic modulus of the metal samples.
    \textbf{b}) The FWHM value of the peak of the second PCA${}_1$ component plotted against the elastic modulus of the metal samples.
    \textbf{c}) The minimum value of the first PCA${}_1$ component plotted against the location of the peak in PCA${}_2$ in the creep tests.
    \textbf{d}) The maximum value of PCA${}_1$ plotted against the location of the peak in PCA${}_2$ in the creep tests.
    \textbf{e}) The range of values of PCA${}_2$ plotted against the location of the peak in PCA${}_2$ in the creep tests.
    \textbf{f}) The maximum value of PCA${}_1$ plotted against the sample lifetime in the creep tests.
    }
    \label{fig:figCorr}
\end{figure}

\subsection{Detection accuracy}

We can compare the yield strains obtained from the stress-strain curves (using Eq.~\ref{eq:yield}) and the ones obtained using the strain fields from DIC, the PCA method and Eq.~\ref{eq:prediction}. Indeed, as one can see from Fig.~\ref{fig:fig4}a, these values match very well. The PCA predicted yield strains are slightly larger than the ones measured from the stress-strain curve as previously seen in simulations \cite{papanikolaou2021direct}. As can be seen in the figure, a line offset by 0.02 \% fits the data very well. The simulation done for Haynes~230 also matches these results despite the higher yield strain and differing curvature around the yield point.\\

Similarly for the creep case, we plot the time of the strain rate minimum determined from the strain rate curve $t_{min}$ versus the value obtained from Eq.~\ref{eq:prediction2} (see Fig.~\ref{fig:fig4}b), and see that they correspond to roughly equal times. There is more scatter on the values obtained using the PCA method but the average is located on the line $t_{min}^{PCA} = t_{min}$ (black line) and agrees with the Monkman--Grant relation $t_{min}/t_c = 0.83$.

One must keep in mind that the Monkman--Grant relation is indeed just a statistical one and additionally the determination of the minimum of a noisy strain rate signal is prone to errors. Instead, the PCA method directly detects the point of maximum fluctuations from the DIC strain fields.

\subsection{Correlation between the PCA components and mechanical properties}
The inference of the yield point and tertiary creep onset are clearly just the tip of the iceberg in the context of extracting material properties from DIC data. As we show in  Fig.~\ref{fig:figCorr}, there seems to be rich information in the PCA behavior that can be associated to a wealth of material properties. As an indication, at this stage, we point out that the maximum of the first PCA component (when not normalized to unity) in the study of tension in Ni-based alloys appears to be proportional to the sample's elastic modulus (cf. Fig.~\ref{fig:figCorr}(a)). In addition, the width of the peak (measured by the full width at half maximum (FWHM) value taken from the middle of the maximum and minimum values) of the second component is also proportional to the sample's elastic modulus. (cf. Fig.~\ref{fig:figCorr}(b)). Then, in the tertiary creep onset investigation, both the minimum and maximum of the first principal component projection (c, d), as well as their mutual difference, (e), are all proportional to the critical ration $t_{min}^{PCA}/t_c$, and finally the maximum of the first PCA component projection is proportional to sample lifetime $t_c$. While these findings are promising, further work, beyond the scope of the current investigation, needs to be pursued to associate these findings to theoretical modeling.

\section{Conclusions}

The results presented here show that this method is effective in determining the yielding of different materials from strain fields obtained through DIC. The behavior was also verified for a simulation done using a crystal plasticity model. The key features exploited in this detection method are the fluctuations in the local strain fields. 

The transition is seen in the PCA components, which characterize the spatial fluctuations. Around the transition point the first component evolves from a low value, which we take to correspond to the elastic state, to a high value, which represents the plasticity dominated state. The second component peaks at the transition point, around the maximal rate of change in the first component.\\

The yield points we determined from the stress-strain curves differ between materials as well as between samples, so using a constant offset yield point (such as 0.2~\% engineering strain) would miss a lot about the details of yielding. In our case, using the criterion of 0.2~\% strain would overestimate the yield strain and therefore also the yield stress.
The point of maximal fluctuations determined by our detection method correlates with the point of maximal curvature of the stress-strain curve, although the fluctuations seem to be maximized slightly (around 0.02 \%) later. The advantage of our method is having a uniquely defined peak instead of constant thresholds or fixed points.\\

Moreover we found that despite of differences in the deformation mechanisms, the same method can be used to detect the onset of tertiary creep in static loading, i.e. the start of acceleration towards failure. The detected point correlates with the time of the strain rate minimum determined from the global strain rate signal and with the statistical Monkman--Grant relation for the location of the minimum, although the scatter detected points is fairly large.
This shows that the spatial fluctuations are a good universal indicator of plasticity and can be exploited in systems with vastly different microstructures.\\

This opens up many practical applications for the method, as it can be used outside of controlled tensile tests. DIC analysis can be applied to any system which can be reliably imaged at different points in time and the PCA method can be used to study the system under for example static loading conditions. We note that no particularly great image resolution is required.

\begin{acknowledgments}
M.J.A. acknowledges support from the Academy of Finland (Center of Excellence program, 278367 and 317464): The authors thank the support from the European Union Horizon 2020 research and innovation programme under grant agreement No 857470 and from European Regional Development Fund via Foundation for Polish Science International Research Agenda PLUS programme grant No MAB PLUS/2018/8.
The authors acknowledge the computational resources provided by the Aalto University School of Science ``Science-IT'' project.
\end{acknowledgments}

%

\end{document}